\documentclass[10pt]{iopart}
\usepackage{iopams}  
\usepackage{graphicx}  
\usepackage{color}
\newcommand\SL{\mathop{\mathrm{SL}}\nolimits}
\newcommand\sly{\SL_y(2)}
\newcommand\slx{\SL_x(2)}
\newcommand\sld{\SL_x(2)\times\SL_y(2)}
\newcommand\rd{{\rm d}}
\newcommand\re{{\rm e}}

\begin{document}

\title[Symmetry preserving discretization of ODE]{Symmetry preserving discretization of ordinary differential equations.
Large symmetry groups and higher order equations.}

 \author{R Campoamor-Stursberg$^{1}$, M A Rodr\'{\i}guez$^{2}$,  P Winternitz$^{3,4}$}
 
 \address{$^1$Instituto de Matem\'atica Interdisciplinar and Dept. de Geometr\'{\i}a y Topolog\'{\i}a, Facultad de Matem\'aticas, Universidad Complutense, 28040-Madrid, Spain}
 \address{$^2$Dept. de F\'{\i}sica Te\'orica II, Facultad de F\'{\i}sicas, Universidad Complutense, 28040-Madrid, Spain}
 \address{$^3$Centre de Recherches Math\'ematiques et D\'epartement de math\'ematiques et de statistique, Universit\'e de Montr\'eal, CP 6128, Succ Ville Marie, H3C 3J7 Montr\'eal (QC), Canada.}
 \address{$^4$On  sabbatical leave at  Dipartimento di Fisica, Universit\`a degli studi Roma Tre, via della Vasca Navale 84, 00146-Roma, Italy.}

\ead{$^1$rutwig@ucm.es, $^2$rodrigue@fis.ucm.es, $^3$wintern@crm.umontreal.ca}
\date{October 30, 2012}
\begin{abstract}
Ordinary differential equations (ODEs) and ordinary difference systems (O$\Delta$Ss) invariant under the actions of the Lie groups $\slx$, $\sly$ and $\sld$ of projective transformations of the independent variables $x$ and dependent variables $y$ are constructed. The ODEs are continuous limits of the O$\Delta$Ss, or conversely, the O$\Delta$Ss are invariant discretizations of the ODEs. The invariant O$\Delta$Ss are used to calculate numerical solutions of the invariant ODEs of order up to five. The solutions of the invariant numerical schemes are compared to numerical solutions obtained by standard Runge-Kutta methods and to exact solutions, when available. The invariant method performs at least as well as standard ones and much better in the vicinity of
singularities of solutions.
\end{abstract}

%\pacs{00.00, 20.00, 42.10}
%\vspace{2pc}
%\noindent{\it Keywords}: 
%\submitto{\JPA}
%\maketitle

\section{Introduction}

The application of Lie groups to the study of difference equations is a relatively new topic that has been actively pursued for the last 30 years or so. For  recent reviews we refer the reader to \cite{Do93,Do01,Do10,Hy14,LO11,LW06,Wi04,Wi09}. Some of the original articles pertinent for this study are \cite{Do91,DK15,DK00a,DK00b,DK04,FV95,LW91,RW09,RV13,RV15}.

This line of research has several aspects. From the point of view of physics, one aim is to preserve such fundamental symmetry properties as Lorentz, Galilei and conformal invariance in a discrete space-time. From the point of view of mathematics, both pure and applied, the aim is to turn Lie group theory into an efficient tool for studying the solution space of difference equations as it has long been for differential ones \cite{BK82,Li88,Ol93,Ov62}. From the point of view of computing, this approach belongs into the field of geometrical integration \cite{HL10,IM00,Is08,QM06}. The aim is to improve the qualitative and quantitative features of numerical solutions of differential equations by introducing difference systems that have the same Lie point symmetry groups as their continuous limits (invariant discretization).

The original idea \cite{Do91} has been applied to both ordinary differential equations (ODEs) \cite{BC06,DK00a,DK00b,DK04,DW00,LT00,RW09,RW04} and partial differential equations (PDEs) \cite{BC15,BD01,DK00a,DW00,LM15a,LM15b,LR13,LR14,LT01,RV13,RV15,VW05}.

For first order ODEs the method provides exact discretizations, i.e. differential systems that have the same solutions as the ODEs \cite{RW04}. For second and higher order ODEs invariant discretization often provides difference schemes that can be solved analytically, using either a Lagrangian \cite{DK00a,DK04} or the adjoint equation method \cite{DK15}. It was shown for numerous second and third order ODEs that invariant discretization provides qualitatively better fits to solutions than standard methods, specially in the neighborhoods of singularities \cite{BC06,RW09}.

The purpose of this article is to extend the method of invariant discretization of differential equations to larger Lie groups and higher order ODEs than have been treated so far. More specifically, we consider the direct product group $\mathrm{SL}_x(2)\times \mathrm{SL}_y(2)$, where $x$ and $y$ are the independent and dependent variables, respectively and treat ODEs up to order 5.

In Section 2 we briefly outline the general method of invariant discretization for ODEs. In Section 3,  we sum up the differential invariants of $\mathrm{SL}_y(2)$ (to all orders), and of $\mathrm{SL}_x(2)$ and $\mathrm{SL}_x(2)\times \mathrm{SL}_y(2)$ (up to order 5). The main results of the article are presented in Section 4. Thus we derive complete sets of difference invariants up to order 5 (using 6 points on a stencil) for all 3 groups under consideration and show how to obtain the differential invariants in the continuous limit. Section 5 is devoted to numerical examples in which we compare results using the invariant discretization with standard numerical methods.

\section{Differential and difference invariants of a Lie group}

Let us consider a Lie group of local point transformations acting on a Euclidean plane with Cartesian coordinates $(x,y)$ generated by a Lie algebra of vector fields of the form
\begin{equation}\label{vector}
X=\xi(x,y)\partial_x+\phi(x,y)\partial_y
\end{equation}

We can view the Lie group as acting either on solutions ($u=f(x)$) of an ODE:
\begin{equation}
E\equiv y^{(N)}-F(x,y,y',y'',\ldots, y^{(N-1)})=0
\end{equation}
or on solutions of an ordinary difference scheme O$\Delta$S
\begin{equation}\label{scheme}
E_a\equiv E_a(\{x_k,y_k\}_{k=0}^{N})=0,\quad a=1,2.
\end{equation}
The difference scheme (\ref{scheme}) that we use consists of two equations, each connecting the $N+1$ points, satisfying
\begin{equation}
\frac{\partial(E_1,E_2)}{\partial (x_N,y_N)}\neq 0
\end{equation}
so that we can calculate $(x_N,y_N)$ if the previous points $(x_k,y_k)$ are known. In the continuous limit we put
\begin{equation}\label{limit}
x_n-x_{n-1}=h_n,\quad h_n=\alpha_n\epsilon,\quad  \epsilon \to 0
\end{equation}
where $\alpha_n$ are some constants of the order $\alpha_n\sim 1$ and we require
\begin{equation}
\{E_1=0\}\to \{E=0\}, \quad \{E_2=0\}\to \{0=0\},\quad \epsilon\to 0
\end{equation}
Thus the lattice equation goes into an identity and the difference scheme goes into the target ODE.

We will take a given group $G$ and find a basis for the differential invariants up to a certain order (i.e. fix the order $N$ of the highest derivative) and for the difference invariants up to the corresponding order (i.e. fix the number of points of the lattice to be used as $N+1$). The ODEs and the O$\Delta$Ss will be constructed out of the respective invariants,  $I_1,\dots, I_K$ or $I^D_1,\ldots, I^D_L$ respectively and we have $K<L$.

Equations written in terms of these invariant will be ``strongly invariant''. Other equations may be ``weakly invariant'', i.e. invariant on some submanifold. We will encounter such cases below for both ODEs and O$\Delta$Ss.

Basically two methods exist for calculating invariants of a given group action on a homogeneous manifold. One is the infinitesimal method based on the prolongations of the vector fields representing the Lie algebra of the group \cite{Li88,Ol93}. The other method is a global one, called the method of moving frames \cite{Ca37,FO98,Ol01}. In the second method it is necessary to express the group parameters in terms of the values of a sufficient number of the transformed variables on some section of the generic orbits. For simple and semisimple groups this typically leads to algebraic equations to solve. In particular for the $\mathrm{SL}_x(2)\times \mathrm{SL}_y(2)$ action studied in this article that leads to a third order algebraic equation. We find the infinitesimal method more convenient for the problem at hand and we use it throughout the article.

The group $G$ and the vector fields $X$ of (\ref{vector}) act on the variables $(x,y)$ and on functions $y=f(x)$ in the same manner, whether we are considering differential equations or difference systems. However in the continuous case we prolong to actions on derivatives in a a standard manner \cite{Ol93}. In the discrete case we write $X$ at some point $x_k$ of the one dimensional lattice and then sum over all points involved in the O$\Delta$S \cite{Do01,LW06}:
\begin{equation}
\mathrm{pr}^{D}X_n=\sum_k\left(\xi(x_{n+k},y_{n+k})\partial_{x_{n+k}}+\phi(x_{n+k},y_{n+k})\partial_{y_{n+k}}\right)
\end{equation}
The summation over $k$ is over all points on one stencil. The index $n$ labels the position of the stencil used in the calculation. 

In both cases we find the invariant by solving the system of determining equations following from the invariance condition
\begin{equation}\label{procont}
\mathrm{pr}X\Phi(x,y,y',\ldots, y^{(N)})=0
\end{equation}
or 
\begin{equation}\label{pres}
 \mathrm{pr}^DX_n\Phi(x_{n+k},y_{n+k})=0
\end{equation}

There will be more functionally independent difference invariants than differential ones. We will divide the difference ones into two sets; those that go into differential ones in the continuous limit and those that vanish in this limit.

The connection between difference and differential invariants is established by using Taylor expansions of the discrete quantities. We restrict ourselves to a single stencil, i.e. points $(x_0,y_0),\ldots, (x_N,y_N)$, and choose a point about which to develop, say $x_0$. All other points are expressed as:
\begin{equation}\label{serx}
x_k=x_0+\sum_{l=1}^k h_l,\quad y_k=y(x_k),\quad 1\le k\le N
\end{equation}
and we expand all discrete invariants using the truncated Taylor series:
\begin{equation}\label{sery}
y_k=\sum_{j= 0}^N\frac{1}{j!} y_0^{(j)} \left(\sum_{l=1}^k h_l\right)^j
\end{equation}
 The continuous limit is taken as in (\ref{limit}). The result will be expressed in terms of the basis of differential invariants. In general the limit may depend on the constants $\alpha_n$ in (\ref{limit}). These will be specific numbers once the lattice is chosen. Detailed examples will be given in Section 4.

\section{Differential invariants under $\mathrm{SL}_y(2)$, $\mathrm{SL}_x(2)$ and $\mathrm{SL}_x(2)\times \mathrm{SL}_y(2)$\label{difftial}}

In this section we restrict the group $G$ to be $\mathrm{SL}_y(2)$, $\mathrm{SL}_x(2)$ and $\mathrm{SL}_x(2)\times \mathrm{SL}_x(2)$ respectively,  and will present bases for all differential invariants up to order 5, though it would be easy to proceed to higher orders. The reason for this choice is that $N=5$ is the lowest order at which an $\mathrm{SL}_x(2)\times \mathrm{SL}_y(2)$ invariant exists.

\subsection{Invariants of $\mathrm{SL}_y(2)$}

The Lie algebra of the group $\mathrm{SL}_y(2)$ is generated  by vector fields $\partial_y$, $y\partial_y$,  $y^2\partial_y$ with prolongations:
\begin{equation}\eqalign{
\mathrm{pr}^{(N)}\partial_y&= \partial_y\\
\mathrm{pr}^{(N)}y\partial_y&= y\partial_y+\sum_{k=1}^N y^{(k)}\partial_{y^{(k)}}\\
\mathrm{pr}^{(N)}y^2\partial_y&= y^2\partial_y+\sum_{k=1}^N  (y^2)^{(k)}\partial_{y^{(k)}},}
\end{equation}
Solving the corresponding PDEs (\ref{procont}) for $N=5$ we find the lowest order differential invariant
\begin{equation}
J_3=\frac{y'''}{y'}-\frac{3}{2}\left(\frac{y''}{y'}\right)^2
\end{equation}
The third and fourth invariants can also be calculated directly.

Alternatively, since the variable $x$ is invariant we can start from the lowest order invariant involving derivatives of $y$, namely $J_3$ and generate a different basis of $\mathrm{SL}_y(2)$ invariants, by invariant differentiation:
\begin{equation}\label{invJ}
J_{k+3}=\frac{\rd^{k} }{\rd x^{k}} J_3,\quad k=1,2,\ldots
\end{equation}
All $\mathrm{SL}_y(2)$ differential invariants of order up to $N$ will be functions of
\begin{equation}
\{x,J_3,J_4,J_5,\ldots ,J_N\}
\end{equation}
We mention that $J_3$ is the well known Schwarzian derivative with many interesting applications \cite{OT14}. We will use in this work the first three invariants, $J_3$, $J_4$ and $J_5$. The explicit form of $J_4$ and $J_5$ are:
\begin{equation}
J_4\equiv J_3'=  \frac{y^{(4)}}{y'}-4 \frac{y''y'''}{y'^2}
  +3 \frac{y''^3}{y'^3} 
\end{equation}
\begin{equation}
J_5\equiv J_3''=\frac{y^{(5)}}{y'}-5\frac{y''y^{(4)} }{y'^2}+17\frac{ y''^2y''' }{y'^3}-4\frac{y'''^2}{y'^2}-9\frac{y''^4}{y'^4}
\end{equation}
We could also use a simplified fifth order invariant, adding a multiple of $J_3^2$
\begin{equation}
\tilde{J}_5\equiv J_5+4J_3^2= \frac{y^{(5)}}{y'}-5 \frac{y''y^{(4)}}{y'^2}+5  \frac{y''^2y'''}{y'^3}
\end{equation}

\subsection{Differential invariants of $\mathrm{SL}_x(2)$}

The Lie algebra of $\mathrm{SL}_x(2)$ has the basis $\partial_x$, $x\partial_x$, $x^2\partial_x$ with prolongations
\begin{equation}\label{prolxx}\eqalign{
\mathrm{pr}^{(N)}\partial_x&=\partial_x\\
\mathrm{pr}^{(N)}x\partial_x&=x\partial_x-\sum_{k=1}^Nky^{(k)}\partial_{y^{(k)}} \\
\mathrm{pr}^{(N)}x^2\partial_x&=x^2\partial_x-\sum_{k=1}^Nk\big((k-1)y^{(k-1)}+2xy^{(k)}\big)\partial_{y^{(k)}} }
\end{equation}
It is of course equivalent to $\mathrm{SL}_y(2)$ and the two are transformed into each other by a hodograph transformation. We treat $\mathrm{SL}_x(2)$ separately here since we are interested mainly in the action of $\mathrm{SL}_x(2)\times \mathrm{SL}_y(2)$.

The three lowest order invariants are (higher order ones are also computed in a straightforward way \cite{Ca04})
\begin{equation}
\eqalign{
K_3=\frac{1}{(y')^2}\left(\frac{y'''}{y'}-\frac{3}{2}\left(\frac{y''}{y'}\right)^2\right)
,\\ 
K_4=\frac{y^{(4)}}{y'^4}-6\frac{y'''y''}{y'^5} +6\frac{y''^3}{y'^6},\\ 
K_5=
\frac{y^{(5)}}{ y'^5}
-10\frac{ y^{(4)}y''}{ y'^{6}}
-4 \frac{y'''^2}{ y'^{6}}
+42\frac{y'''y''^2}{ y'^{7}}
-\frac{63}{2}\frac{y''^4}{y'^8}}
\end{equation}
and, of course, $y$ itself is an invariant.

It is interesting to consider the behavior  of the $\mathrm{SL}_y(2)$ invariant $J_3$. We have
\begin{equation}
(\mathrm{pr}^{(3)}\partial_x)J_3 =0,\quad (\mathrm{pr}^{(3)} x\partial_x )J_3=-2J_3,\quad (\mathrm{pr}^{(3)}  x^2\partial_x )J_3=-4x J_3
\end{equation}

Thus, while $J_3$ is not invariant under $\mathrm{SL}_x(2)$, the equation 
\begin{equation}\label{weak}
J_3=0
\end{equation}
determines an invariant manifold and the equation (\ref{weak}) is ``weakly invariant'' under the entire group $\mathrm{SL}_x(2)\times \mathrm{SL}_y(2)$. The same of course holds for $K_3=0$.

\subsection{Differential invariants of $\mathrm{SL}_x(2)\times \mathrm{SL}_y(2)$}

We start from the invariants of $\mathrm{SL}_y(2)$, $I(J_3,J_4,J_5,)$ and require that this function be annihilated by the vector fields (\ref{prolxx}). We find that there are no differential invariants of order $N<5$ and only one of order 5, namely
\begin{eqnarray}\label{invH}
\fl H_5=\frac{J_5}{J_3^2}-\frac{5J_4^2}{4J_3^3} Ê=\frac{K_5}{K_3^2}-\frac{5K_4^2}{4K_3^3} \nonumber \\=\frac{2}{(2 y''' y'-3 y''^2)^3}  \bigg(2 y'^3  (2  y'y'''-3 y''^2 )y^{(5)} +20 y'^3 y''y''' y^{(4)} -5y'^4 (y^{(4)})^2 \nonumber \\  -16  y'^3y'''^3+12 y'^2 y''^2y'''^2 -18  y' y''^4y''' + 9 y''^6\bigg)
\end{eqnarray}

\section{Difference invariants and their continuous limits for the groups $\mathrm{SL}_x(2)$, $\mathrm{SL}_y(2)$ and $\mathrm{SL}_x(2)\times \mathrm{SL}_y(2)$}

\subsection{General comments. The cross-ratios}

The Lie group actions that we are considering in this article are the standard projective action of $\mathrm{SL}(2)$ on a real or complex line (the action of the M\"obius group). The fundamental invariants of this action are well-known (and can easily be reobtained using the prescription (\ref{pres})). The lowest order invariants involve 4 points and are the cross-ratios (anharmonic ratios):
\begin{eqnarray}
R_{k+3} = \frac{(y_{k+3}-y_{k+1})(y_{k+2}-y_k)}{(y_{k+3}-y_{k+2})(y_{k+1}-y_k)}\label{invR}\\
S_{k+3} = \frac{x_{k+3}-x_{k+1})(x_{k+2}-x_k)}{(x_{k+3}-x_{k+2})(x_{k+1}-x_k)}\label{invS}
\end{eqnarray}
 for $\mathrm{SL}_y(2)$ and $\mathrm{SL}_x(2)$ respectively.

For $\mathrm{SL}_x(2)\times \mathrm{SL}_y(2)$ these are the only four-points invariants and all higher order invariants can be formed by shifting the four points to the right, forming e.g. the cross-ratios $R_{k+4}$,$R_{k+5}$ \dots, and taking linear combinations of $R_{k+3}$, $R_{k+4}$ \dots, etc. The problem is to form the linear combinations that will have the chosen differential invariants of Section \ref{difftial} as a continuous limit.

For $\mathrm{SL}_x(2)$ and $\mathrm{SL}_y(2)$ we have further difference invariants, namely the dependent variables $y_n$ and the independent variables $x_n$ for $\mathrm{SL}_x(2)$ and $\mathrm{SL}_y(2)$,  respectively

The cross-ratios $S_j$ will be used to write invariant lattices, e.g. $S_{k+3}=A$ or $S_{k+3}=AS_{k+4}$, where $A$ is a constant.

The cross-ratios $R_j$ will be expanded into power series, and using equations (\ref{serx}) and (\ref{sery}) we relate them to differential invariants.

Let us consider $\mathrm{SL}_y(2)$, $\mathrm{SL}_x(2)$ and $\mathrm{SL}_x(2)\times \mathrm{SL}_y(2)$ separately.

\subsection{The group $\mathrm{SL}_y(2)$}

Taking 4 adjacent points, $x_0,x_1,x_2,x_3$ (we simplify the notation writing $x_{n+k}\equiv x_k$) and expanding around $x_0$ we obtain:
\begin{eqnarray}  \fl
R_{3} = S_{3}\bigg[1-\frac{1}{6} h_{2} (h_{1}+h_{k+2}+h_{k+3}) \bigg(J_3 
+\frac{1}{4} (3 h_{1}+ 2h_{k+2}+h_{3}) J_4\nonumber  \\
+\frac{1}{60}\bigg(3(6 h_{1}^2+3 h_{2}^2+h_{3}^2+ 8h_{1} h_{2}+4h_{1}h_{3} +3 h_{2} h_{3} ) J_5\nonumber \\
-10(7h_{1}^2+4 h_{2}^2+h_{3}^2+10 h_{1}h_{2}+5h_{1}h_{3} +4 h_{2} h_{3} ) J_3^2\bigg)\bigg)\bigg] \nonumber \\ +O(h^5)\label{eqR}
\end{eqnarray}

We can make a similar expansion for two more sets of four points, $x_1,x_2,x_3,x_4$ and $x_2,x_3,x_4,x_5$ (always expanding around $x_0$). We see that to lowest order in $h$ the difference invariant that has the correct continuous limit is
\begin{equation}\label{invL3}
L_{k+3}^{(3)}=\frac{6}{(x_{k+2}-x_{k+1})(x_{k+3}-x_k)}\left(1-\frac{R_{k+3}}{S_{k+3}}\right);
\end{equation}
indeed for all values of $k$ (we will only need $k=0,1,2$) we have
\begin{equation}
\lim_{h_j\to 0}L^{(3)}_{k+3}=J_3.
\end{equation}

In view of (\ref{eqR}) we can define a new set of difference invariants:
\begin{equation}\label{invL4}
L^{(4)}_{k+4}=\frac{4}{x_{k+4}-x_k}\left(L^{(3)}_{k+4}-L^{(3)}_{k+3}\right)
\end{equation}
which can be expanded in $h$ as:
\begin{eqnarray}
\fl L^{(4)}_{4}=J_4+\frac{1}{15 (h_{1}+h_{2}+h_{3}+h_{4})}  \bigg[3 \bigg(4 h_{1}^2+3 h_{2}^2+2 h_{3}^2+h_{4}^2\nonumber \\  +7 h_{1}h_{2}+6 h_{1}h_{3}+5 h_{1}h_{4} +5 h_{2} h_{3}+4 h_{2} h_{4}+3 h_{3} h_{4}\bigg)  J_5 \nonumber \\  -2 \left(h_{1}^2-3 h_{2}^2+3 h_{3}^2-h_{4}^2-2 h_{1}h_{2}-h_{1}h_{3} +h_{2} h_{4}+2 h_{3} h_{4}\right) J_3^2\bigg]\nonumber \\ +O(h^2)
\end{eqnarray}
and a similar expression for $L^{(4)}_5$. In the continuous limit, we have, for all $k$ (in particular, for the only two values we need, $k=0,1$):
\begin{equation}\label{j4}
\lim_{h_j\to 0} L^{(4)}_{k+4}=J_4
\end{equation}

Similarly, to obtain the fifth order differential invariant in the continuous limit we form another set of difference invariants, namely 
\begin{equation}\label{invL5}
L^{(5)}_{5}=\frac{5}{x_{5}-x_{0}}\left(L^{(4)}_{5}-L^{(4)}_{4}\right)
\end{equation}
and
\begin{eqnarray}
\fl L^{(5)}_{5}=J_5+ \frac{10}{3} \bigg[\frac15-\frac{h_{4}}{h_{1}+h_{2}+h_{3}+h_{4}+h_{5} }\nonumber \\Ê
 -\frac{ (h_{1}+h_{2}) (h_{2}+h_{3})}{  (h_{1}+h_{2}+h_{3}+h_{4}+h_{5}) (h_{1}+h_{2}+h_{3}+h_{4})}\nonumber \\ 
 +\frac{  (h_{2}+h_{3}) (h_{3}+h_{4})}{  (h_{1}+h_{2}+h_{3}+h_{4}+h_{5})(h_{2}+h_{3}+h_{4}+h_{5})}\bigg] J_3^2+O(h).
\end{eqnarray}
From (\ref{eqR}) (e.g. for $k=0$, 1 and 2) we see that the limit of $L^{(5)}_{5}$ is not exactly $J_5$ of (\ref{invJ}) but rather a combination of $J_5$ and $J_3^2$
 \begin{equation}
 \lim_{h_j\to 0}L^{(5)}_{5}=J_5+W_0J_3^2
 \end{equation}
 with 
 \begin{equation}
 W_0=\lim_{h_j\to 0}W
 \end{equation}
 and
 \begin{equation}\label{wexp}
 \fl W=\frac{10}{3}\left(\frac15-\frac{x_{4}-x_{3}}{x_{5}-x_{0}}-\frac{(x_{3}-x_{1}) (x_{2}-x_{0})}{(x_{5}-x_{0})(x_{4}-x_{0})}+\frac{(x_{4}-x_{2}) (x_{3}-x_{1})}{(x_{5}-x_{0})(x_{5}-x_{1})}\right)
 \end{equation}
The coefficient $W$ depends on the specific form of the lattice, and $W_0$ is a finite number (not necessarily zero).

For instance, let us consider an $\mathrm{SL}_x(2)\times \mathrm{SL}_y(2)$ invariant lattice given by
\begin{equation}\label{eqS}
S_j=K,\quad \forall j
\end{equation}
where $K$ is a constant. This equation (\ref{eqS}) was solved in \cite{DK15} for arbitrary values of $K$. The result is particularly simple for $K=4$. A particular solution, not contained in the general one is
\begin{equation}\label{sol1}
x_m=Am+B
\end{equation}
and the general one is
\begin{equation}\label{sol2}
x_m=\frac{1}{Am+B}+C
\end{equation}
($A$, $B$, $C$ are constants). Solution (\ref{sol1}) corresponds to a uniform lattice with $h_m=A$ for all $m$.  For (\ref{sol1}) and (\ref{sol2}) we have   
\begin{equation}
W_0=0,\quad W_0=\frac{2A^2(8A^2-5AB-B^2)}{(A+B)(2A+B)(3A+B)(4A+B)}
\end{equation}
respectively.

Thus, $W_0$ is a definite number (and can be set equal to zero in the case (\ref{sol2}) by choosing $B=\frac12(A(-5\pm \sqrt{57})$).

We see that it is not difficult to construct $\mathrm{SL}_y(2)$ difference invariants of arbitrary orders. The challenge is to find an appropriate basis for these invariants that in the continuous limit reproduces the chosen basis of difference invariant.

The connection between the difference and differential $\mathrm{SL}_y(2)$ invariants of order 3, 4 and 5 is given by equations (\ref{invL3}), (\ref{invL4}) and (\ref{invL5}), respectively.

\subsection{The group $\mathrm{SL}_x(2)$}

The approach used for $\mathrm{SL}_y(2)$ must be modified since the differences $h_k$ are not invariant under $\mathrm{SL}_x(2)$. Instead of (\ref{invL3}) we expand the $\mathrm{SL}_x(2)$ invariants:
\begin{equation}
M^{(3)}_{k+3}=\frac{6}{(y_{k+3}-y_k)(y_{k+2}-y_{k+1})}\left(1-\frac{R_{k+3}}{S_{k+3}}\right)
\end{equation}
and obtain
\begin{eqnarray}
\fl M^{(3)}_{3}=  K_3 
+\frac{1}{4} (3 h_{1}+ 2h_{2}+h_{3}) y'K_4\nonumber \\
  +\frac{1}{120} \bigg(
6 (6 h_{1}^2+3 h_{2}^2+h_{3}^2+8 h_{1}h_{2}+4 h_{1}h_{k+3} +3 h_{2} h_{3} )  y'^2K_5  \nonumber \\
 -4 (19 h_{1}^2+12 h_{2}^2+4h_{3}^2+27 h_{1}h_{2}+11 h_{1}h_{3} +12 h_{2} h_{3} )  y'^2K_3^2  
\nonumber \\Ê
 +15 (3 h_{1}^2+2 h_{2}^2+h_{3}^2+4 h_{1}h_{2}+2 h_{1}h_{3} +2 h_{2} h_{3} )  y''K_4 
 \bigg)  +O(h^3) 
\end{eqnarray}
and similar expressions for $k=1,2$. Thus we have
\begin{equation}
\lim_{h_j\to 0}M^{(3)}_{k+3}=K_3,\quad \forall k
\end{equation}
In analogy, we define
\begin{equation}
M^{(4)}_{k+4}=\frac{4}{y_{k+4}-y_k}\left(M^{(3)}_{k+4}-M^{(3)}_{k+3}\right),\quad \forall k
\end{equation}
satisfying
\begin{equation}
\lim_{h_j\to 0} M^{(4)}_{k+4}=K_4,\quad \forall k
\end{equation}

Finally, to obtain a fifth order differential invariant in the continuous limit we define
\begin{equation}
M^{(5)}_{5}=\frac{5}{y_{5}-y_0}\left(M^{(4)}_{5}-M^{(4)}_{4}\right)
\end{equation}
satisfying
\begin{equation}
\lim_{h_j\to 0}M^{(5)}_{5} =K_5-W_{x,0}K_3^2,\quad W_{x,0}=\lim_{h_j\to 0}W_x
\end{equation}
where \begin{equation}
W_x=\frac{20}{6} \left(\frac45 -\frac{h_3}{ h_1+h_2+h_3+h_4+h_5 }\right)
\end{equation}

As in the case of $\mathrm{SL}_x(2)$ we do  not obtain $K_5$ in the limit but have an additional term involving a scalar multiple, $W_{x,0}$, of a lower order invariant $K_3$. The number $W_{x,0}$ can be evaluated on any lattice and is 
\begin{equation}
W_{x,0}=2
\end{equation}
 on a uniform lattice.

\subsection{The group $\mathrm{SL}_x(2)\times\mathrm{SL}_y(2)$}

For the group $\mathrm{SL}_x(2)\times \mathrm{SL}_y(2)$ all invariants must be constructed out of the cross-ratios $R_{k+3}$, $S_{k+3}$. To obtain the lowest order differential invariant $H_5$ given by (\ref{invH}) we need three values of $k$, e.g. $k=0,1,2$. As always, the difficulty is to identify the combination or combinations of them that go to $H_5$ in the continuous limit. One way to do this is to expand $Q_{k+3}=1-\frac{R_{k+3}}{S_{k+3}}$ for $k=0,1,2$ into power series, and eliminate terms of order $h^2$ and $h^3$ using $S_{k-3}$ whenever possible. Another way is to inspire oneself by the continuous limit (\ref{invH}) and build up an invariant with the correct limit using discretized versions of $J_5$, $J_4$, $J_3$ and $K_5$, $K_4$ $K_3$. Both  methods are quite laborious, even using computer algebra, and lead to the result
\begin{equation}
\mathsf{H}^{(5)} =\frac{(x_{5}-x_{4})(x_{1}-x_{0})}{(x_{4}-x_{3})(x_{2}-x_{1})}
\frac{1}{L^{(3)}_{5}
} \left(\frac{ L^{(5)}}{L^{(3)}_{3}}-\frac{5}{4}
\frac{L^{(4)}_{4}L^{(4)}_{5}}{L^{(3)}_{3}L^{(3)}_{4}}\right)
\end{equation}
or explicitly in terms of the invariants $R_i$ and $S_i$
\begin{eqnarray} 
\fl\mathsf{H}^{(5)}=\frac{10}{3}\frac{S_4}{(S_3 (1-S_4)+S_4) (S_4 (1-S_5)+S_5) (S_4(1-S_3) (1-S_5)-S_3 S_5)}\nonumber \\  \times\left(S_4(1-S_5)\frac{1}{Q_5}  +S_4(1-S_3)\frac{1}{Q_3}\right.\nonumber \\ \left.-(1-S_4) (S_4(1-S_3) (1-S_5)-S_3 S_5)\frac{1}{Q_4} \right. \nonumber \\ \left. -S_4(1-S_3) (1-S_5)\frac{Q_4}{Q_3Q_5}\right)
\end{eqnarray}
where
\begin{equation}
Q_i=1-\frac{R_i}{S_i},\quad i=3,4,5
\end{equation}

The continuous limit is given by (the quantity $W$ was defined in (\ref{wexp})): 
\begin{equation}
\lim_{h_j\to 0}\mathsf{H}^{(5)} =\left(\lim_{h_j\to 0}\frac{h_5h_1}{h_4h_2}\right)(H_5+\lim_{h_j\to 0}W)
\end{equation}

For the case of a uniform lattice ($S_j=4$, $h_j=h_{j+1}$) we have $W=0$ and 
\begin{equation}
\lim_{h\to 0}\mathsf{H}^{(5)} =H_5=\frac{J_5}{J_3^2}-\frac{5J_4^2}{4J_3^3}
\end{equation}
 with
 \begin{equation}
\mathsf{H}^{(5)}=\frac{16 R_5+R_4(3R_4+R_5-32) +R_3 (R_4-5 R_5+16)}{2 (R_3-4) (R_4-4) (R_5-4)}
\end{equation}
For other invariant lattices we have $W=$ constant $\neq 0$.

The overall conclusion from this section is that we have constructed all difference and differential invariants up to order 5 for all groups considered. We have also shown how to proceed to higher orders, using the fundamental invariants, specifically the cross-ratios.

We now proceed to test the invariant O$\Delta$Ss presented in this section and numerical schemes for specific equations. So far, only the group $\sly$ has been used in this manner and only for ODEs of order 2 or 3 \cite{BC06,RW09}.

\section{Numerical examples}

We present in this section some representative examples of differential equations which are invariant under the groups $\mathrm{SL}_y(2)$, $\mathrm{SL}_x(2)$ and $\mathrm{SL}_x(2)\times \mathrm{SL}_y(2)$. Some particular solutions are numerically computed using the invariant discretizations studied in this work. In some cases the discretized scheme provides the exact solution. In other cases, the roundoff errors will forbid to find the solutions at some points. Increasing the working precision would allow to find more approximate solutions, but we cannot keep the same working precision beyond a certain point. These quantitative aspects of the theory have been studied elsewhere \cite{BC06} and will not be discussed in detail in this work. Since our purpose is to provide some qualitative remarks on the invariant method, the standard numerical approach has been carried out using the software Mathematica, which allows a high level performance in all the examples we will discuss below, while being very simple to use. In each case, the program chooses the most appropriate method (for instance, an Adams predictor-corrector or an explicit Runge-Kutta method) and, whenever necessary, a variable step, looking for higher accuracy. In both approaches (standard and invariant) some control parameters, like working precision and accuracy, have also been chosen in order to get the best results.

\subsection{Example 1: fourth order equation, invariant under $\mathrm{SL}_y(2)$}
The differential equation: 
\begin{equation}\label{eq1}
 \frac{y^{(4)}}{y'}-4 \frac{y''y'''}{y'^2}
  +3 \frac{y''^3}{y'^3}=\cos x
\end{equation}
is invariant under $\mathrm{SL}_y(2)$, but not under any transformation of $\mathrm{SL}_x(2)$. Note that the equation can be written as:
\begin{equation}
 J_3'=\cos x
\end{equation}
and a first integral is:
\begin{equation}
 J_3=\sin x+ A
\end{equation}
However, to check the invariant method, we will use the invariant scheme for fourth order differential equations (\ref{j4}). 

We choose as initial data:
\begin{equation}
y(1)=1.0,\quad y'(1)=-1.0,\quad y''(1)=-2.5,\quad y'''(1)=5.0
\end{equation}

\begin{figure}[ht]
\begin{center}
\includegraphics[scale=0.5]{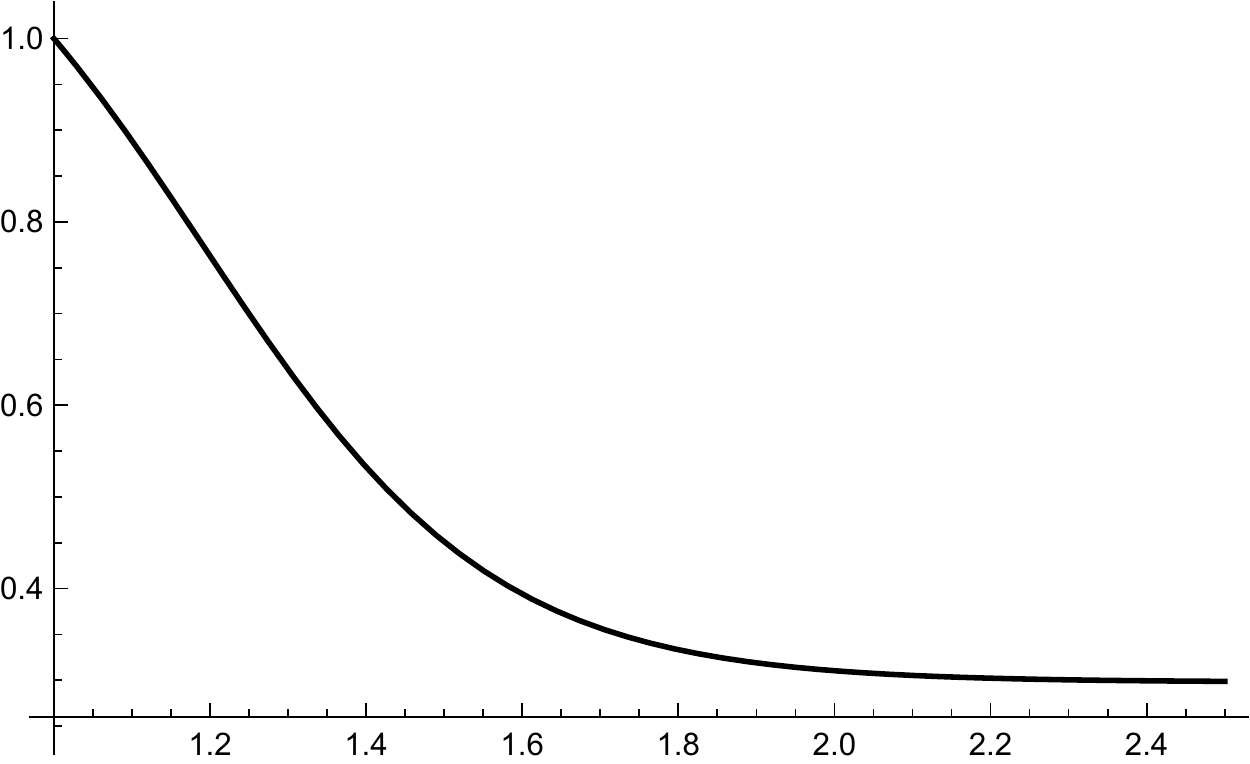}
\caption{Example 1. Solution of equation (\ref{eq1}). Standard numerical method.\label{fig1}}
\end{center}
\end{figure}

Using a standard numerical solution in the interval $[1,2.5]$, we  obtain the graphics of Figure \ref{fig1}. 

We use a uniform lattice with step $h$ (which like any lattice depending only on the variables $x_i$ is invariant under $\mathrm{SL}_y(2)$) on a five-point stencil. The equations are
\begin{equation}
\eqalign{\fl y_n=\left({1-\frac{y_{n-3}-y_{n-2}}{y_{n-3}-y_{n-1}} \left(\frac{(y_{n-1}-y_{n-3}) (y_{n-2}-y_{n-4})}{(y_{n-1}-y_{n-2}) (y_{n-3}-y_{n-4})}-2 h^3 \cos (h (n-2)+x_0)\right)}\right)^{-1} \nonumber \\
\fl  \times \left(y_{n-2}-\frac{y_{n-1} (y_{n-3}-y_{n-2})}{y_{n-3}-y_{n-1}} \left(\frac{(y_{n-1}-y_{n-3}) (y_{n-2}-y_{n-4})}{(y_{n-1}-y_{n-2}) (y_{n-3}-y_{n-4})}\right.\right.\nonumber \\Ê\left.\left.-2 h^3 \cos (x_0+h (n-2))\right)\right)\nonumber \\
 x_n=x_0+nh}
\end{equation}
with initial conditions taken from the numerical solution computed with a standard procedure. We compute the cosine function in the middle point of the stencil and consider three cases, with three different steps $h$.

Table \ref{tab1} compares the values at several points using standard numerical methods (explicit Runge-Kutta) and the invariant approach (Inv) with different steps. The accuracy of the method improves when $h$ goes  to 0, (although the increasing number of steps will correspondingly increase the roundoff errors. We have used the same working precision in all cases).
\begin{table}[h]
\begin{center}
\begin{tabular}{rrrrr}
$x$ & Standard & Inv $h=0.1$ & Inv $h=0.01$ & Inv $h=0.001$ \\
\hline
 1.5 & 0.451089 & 0.451095  & 0.451084  & 0.451089\\
 2.0 & 0.310434 & 0.310466  & 0.310435  & 0.310434\\
 2.5 & 0.298849 & 0.298885  & 0.298850  & 0.298849\\
 \hline
\end{tabular}
\caption{Equation (\ref{eq1}). Values of the solution at points $1.5$, $2.0$ and $2.5$. The columns are the values obtained with a standard numerical procedure and with the invariant one with different steps, respectively. \label{tab1}}
\end{center}
\end{table}

To test the precision of the results we can use the distance, as mean square averages,  between the numerical solutions computed by the invariant scheme and the standard method, defining a global estimator $\chi$:
\begin{equation}\label{chi}
\chi=\sqrt{\frac{\sum_{n}(y^{\mathrm{Inv}}_{n}-y_{n})^2}{\sum_{n}y_{n}^2}}
\end{equation}
and get the results in Table \ref{tab2} at different values of the step, in the interval $[0,2.5]$. The table shows the deviation of the invariant solution from the standard one. The precision of the method is improved when $h$ diminishes, as it was expected. In other cases when we know the exact solution, we could compare $y_n^{\mathrm{Inv}}$  with the exact solution $y_n^{\mathrm{Exact}}$ in (\ref{chi}).

\begin{table}[h]
\begin{center}
\begin{tabular}{cccc}
 $h$ & $ 0.1$ & $0.01$ & $0.001$ \\ 
 \hline
$\chi$ & $ 4.6729.\times 10^{-5}$ & $1.7866\times 10^{-6}$ & $7.2291\times 10^{-8}$\\
\end{tabular}
\caption{Equation (\ref{eq1}). Values of $\chi$. Invariant approach with different steps versus standard numerical approach \label{tab2}}
\end{center}
\end{table}

\begin{figure}[h]
\begin{center}
\includegraphics[scale=0.5]{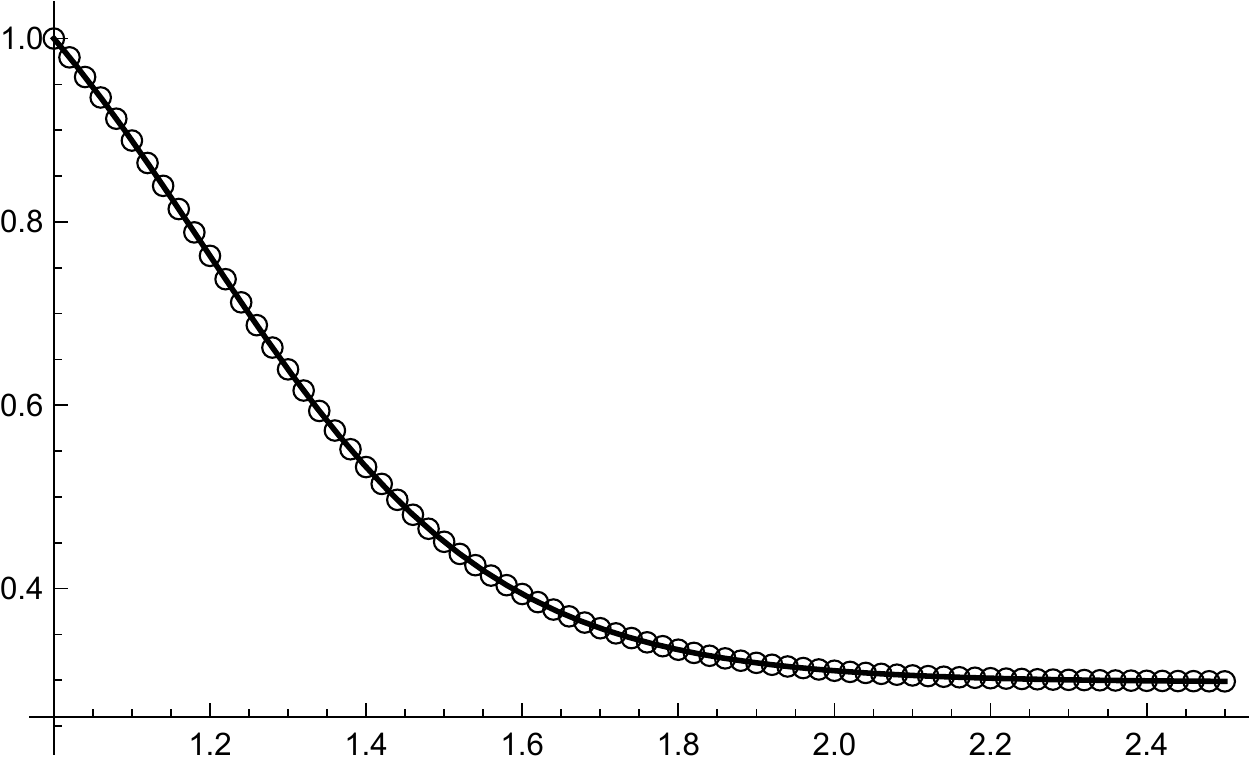}
\caption{Equation (\ref{eq1}). Invariant approach versus standard approach for step $h=0.001$\label{fig2}}
\end{center}
\end{figure}

In Figure \ref{fig2}, the dots are the values of the numerical solutions computed with an invariant approach, for $h=0.001$. The solid curve is the standard numerical approach.

%%%%%%%%%%%%%%%%%%%%%%%%%%%%%%%%%%%%%%%%%%%%%%%%%%%%%%%%%%%%%%%%%%%%%%%%%%%%%%%%%
%%%%%%%%%%%%%%%%%%%%%%%%%%%%%%%%%%%%%%%%%%%%%%%%%%%%%%%%%%%%%%%%%%%%%%%%%%%%%%%%%
%%%%%%%%%%%%%%%%%%%%%%%%%%%%%%%%%%%%%%%%%%%%%%%%%%%%%%%%%%%%%%%%%%%%%%%%%%%%%%%%%

\subsection{Example 2: third order equation invariant under $\mathrm{SL}_x(2)$}

Let us consider the differential equation: 
\begin{equation}\label{eq2}
\frac{1}{y'^2}\left(\frac{y'''}{y'}-\frac{3y''^2}{2y'^2}\right)=c, \quad c\neq 0
\end{equation}
where $c$ is a constant. This equation is invariant under $\mathrm{SL}_x(2)$ (and under translations in $y$). The order can be easily reduced by one, and the general solution in terms of elementary functions can be written as
\begin{equation}\label{arc}
y(x)=c_3+\sqrt{\frac{2}{c}}\mathrm{arctanh}\left(c_1x+c_2\right)
\end{equation}
A 2-parameter family of particular solutions not contained in (62) is
\begin{equation}
y=a+\frac{1}{\sqrt{2c}}\log(x-b)
\end{equation}

When $c=1/2$, a particular solution is:
\begin{equation}\label{log}
y(x)=\log |x|
\end{equation}
which has a singularity at $x=0$. The initial values for the solution in the region $x<0$ can be chosen as
\begin{equation}
y(-1)=0,\quad y'(-1)=-1,\quad y''(-1)=-1
\end{equation}
and the solution can be numerically computed by any standard method which will obviously stop at $t=0$.

We shall now apply the method of invariant discretization to calculate solutions of the type (\ref{arc}) and (\ref{log}) numerically. Let us start with solution (\ref{log}). 
 The invariant approach is obtained with the difference and $x$-lattice equations (we choose a uniform lattice, yielding $S=4$)
\begin{equation}
\eqalign{
\fl \frac{6}{(y_{n-1}-y_{n-2}) (y_{n}-y_{n-3})} \left(1-\frac{(y_{n-3}-y_{n-1}) (y_{n-2}-y_{n})}{4 (y_{n-3}-y_{n-2}) (y_{n-1}-y_{n})}\right)=\frac12,\label{sec}\\
\fl x_n=x_0+ n h}
\end{equation}
The initial conditions are taken from the exact solution:
\begin{eqnarray*}
\fl x_0=-1,\; h=0.0001, \; y_0=0.,
\; y_1=\log(0.9999), \; y_2=\log(0.9998)
\end{eqnarray*}

\begin{figure}[h]
\begin{center}
\includegraphics[scale=0.5]{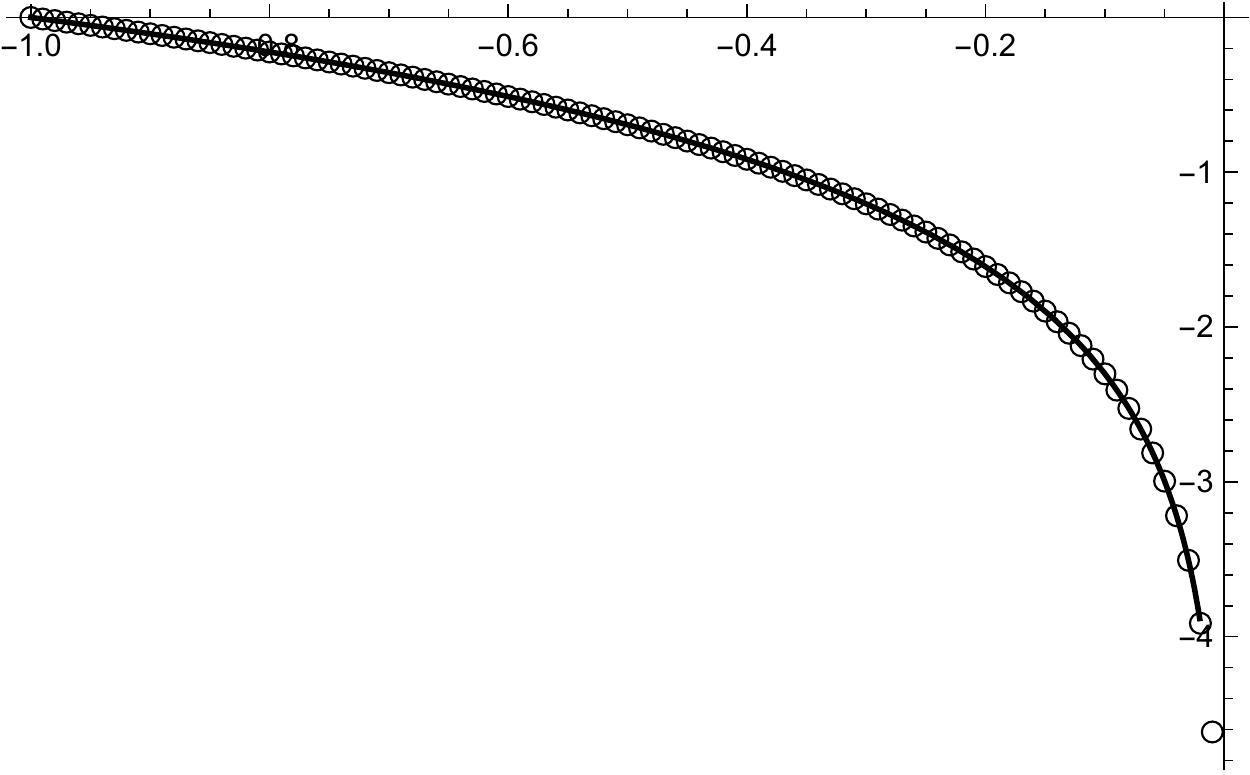}
\caption{Differential equation (\ref{eq2}) $x<0$. Invariant approach, $h=0.0001$, (dots) versus exact solution (solid line).\label{fig3b}}
\end{center}
\end{figure}

However, in contrast with Example 1, the difference equation is not linear in $y_n$, but quadratic. Apart from technical difficulties, one obtains two possible solutions. One of them provides the approximate solution. Note that the difference equation could be, in principle, computed beyond the stop point $x=0$ but $y_n$ becomes complex   if  $x_n$ is greater than 0. See Figure \ref{fig3b} for a graphics of the approximate solution (dots) versus the exact one (solid line).

\begin{figure}[h]
\begin{center}
\includegraphics[scale=0.5]{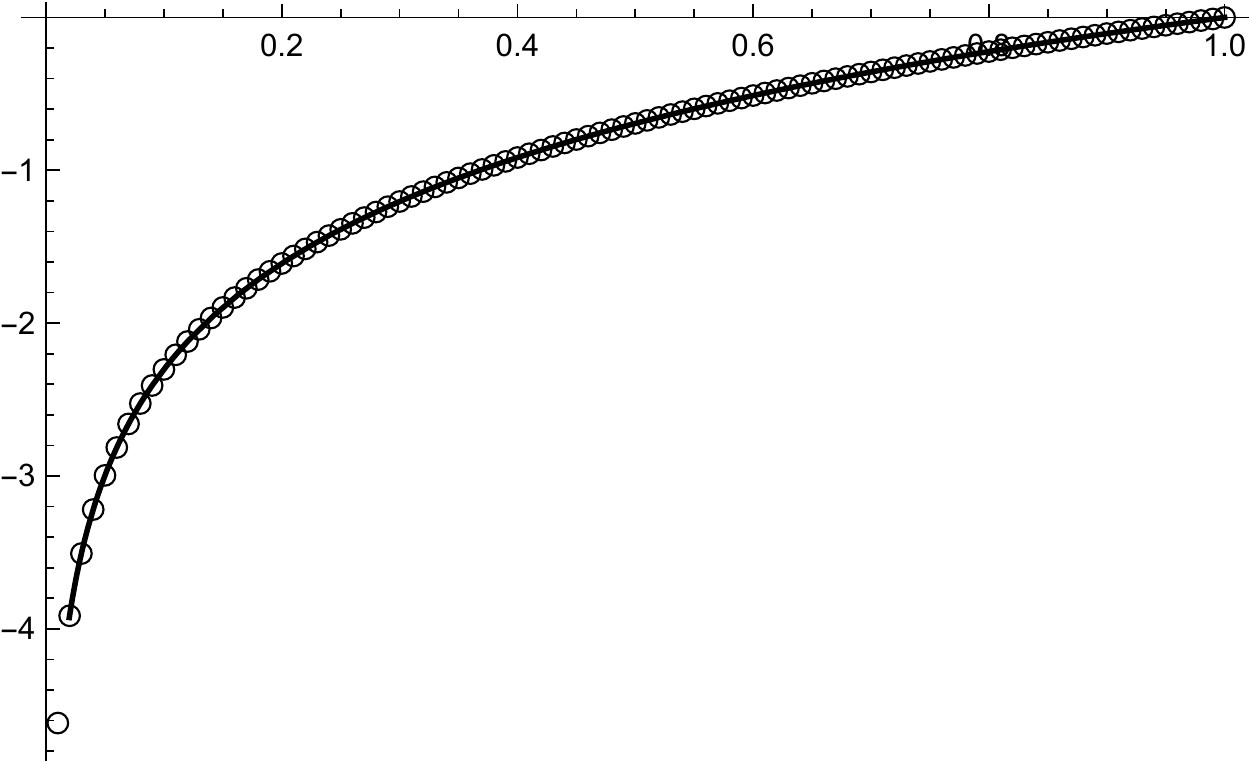}
\caption{Differential equation (\ref{eq2}), $x>0$. Invariant approach, $h=-0.0001$, (dots) versus exact solution (solid line).\label{fig3c}}
\end{center}
\end{figure}

We can also compute the solution in the region $x>0$. Starting at $x_0=1$ and using a negative step $h=-0.0001$, we get for the invariant numerical solution the graphics in Figure \ref{fig3c}. As in the previous case, the $y_n$ values become complex when $x_n$ becomes negative. Then, we have reproduce in these graphics, \ref{fig3b} and \ref{fig3c}, the two regions of the logarithmic solution.

%If we consider the second solution of the difference equation (\ref{sec}) the results do not correspond to the $\log|x|$ solution and cannot be easily interpreted. In fact, a relevant question is how to choose, in the discrete case, the right solution, that which fits the corresponding solution in the continuous case. If we do not know the exact solution we cannot compare and do not have a criteria to select it. However in all the cases we have examined, there is always a solution with values which are close to the initial data, while the other solution, the second root of the quadratic equation, provides values which are much greater than the initial data and should be rejected under the assumption of the existence of smooth solutions. This observation does not close the discussion of this problem. In fact, work is in progress to understand this situation both from the chosen invariant procedure of discretization and also from the symmetry involved in the whole scheme. 

%%%%%%%%%%%%%%%%%%%%%%%%%%%%%%%%%%%%%%%%%%%%%%%%%%%

\begin{figure}[h]
\begin{center}
\includegraphics[scale=0.5]{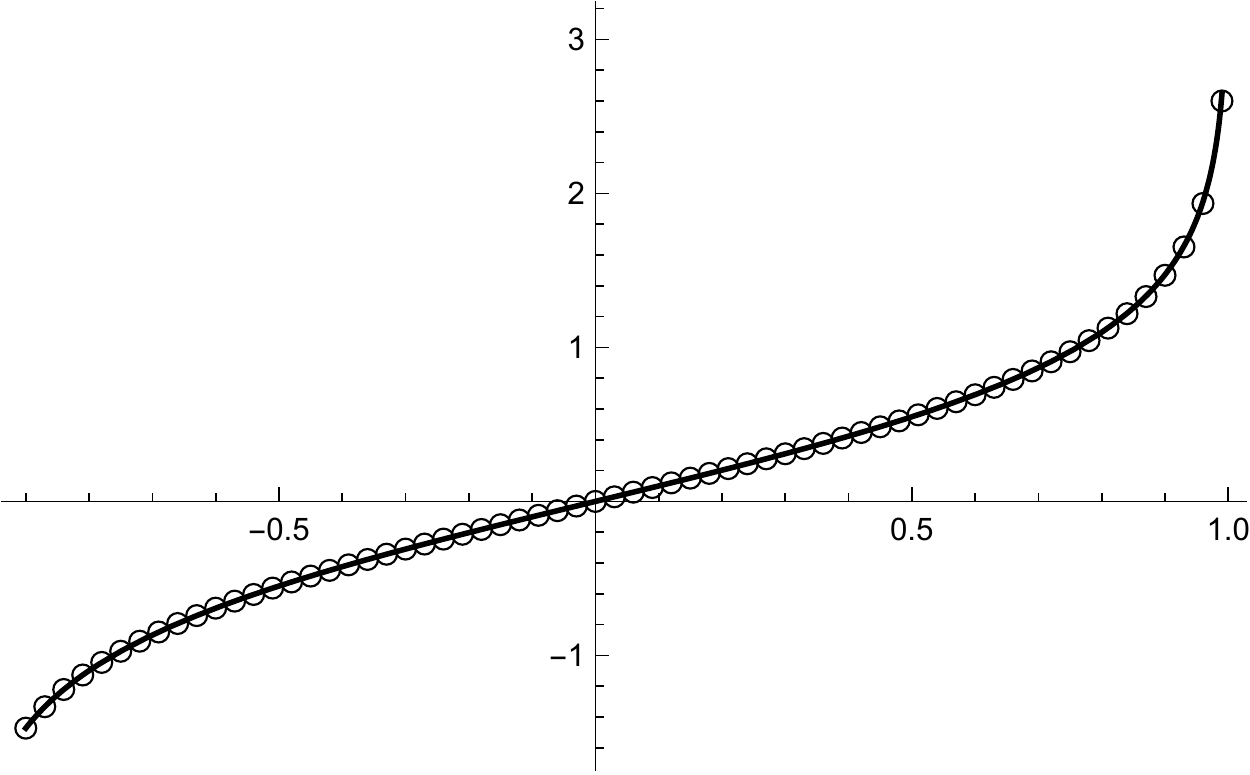}
\caption{Differential equation (\ref{eq2}). Invariant approach, $h=0.001$, (dots) versus exact solution (solid line).\label{fig3}}
\end{center}
\end{figure}

Let us now choose a different solution of (\ref{eq2}) (with $c=2$), namely
\begin{equation}
y(x)=\mathrm{arctanh} \,x
\end{equation}
which exists only in the interval $(-1,1)$.  The initial values for this solution are:
\begin{equation}
y(0)=0,\quad y'(0)=1,\quad y''(0)=0
\end{equation}
and can be numerically computed by any standard method which will stop at $x=1$.
The invariant approach is obtained with the same difference equation as above (\ref{sec}), but the initial conditions (which are again taken from the exact solution) are:
\begin{eqnarray*}
\fl x_0=-0.9,\; h=0.01, \; y_0=\arctan(-0.9),
\; y_1=\arctan(-0.89), \; y_2=\arctan(-0.88)
\end{eqnarray*}
As above, one of the two solutions for $y_n$ provides the approximate solution. See Figure \ref{fig3} for a plot of the invariant numerical solution (dots) versus the exact one (solid line).
The $\chi$ estimator in the interval $[-0.9,0.9]$ can be computed against the exact solution, for different values of the step, see Table \ref{tab3}. 
\begin{table}[h]
\begin{center}
\begin{tabular}{cccc}
$h$ & $0.1$ & $0.01$ & $0.001$ \\ 
\hline
$\chi$ & $0.145901$ & $ 0.007028$  & $0.001131$\\
 \end{tabular}
\caption{Equation (\ref{eq2}). $\chi$ for different values of the step $h$.\label{tab3}}
\end{center}
\end{table}

%%%%%%%%%%%%%%%%%%%%%%%%%%%%%%%%%%%%%%%%%%%%%%%%%%%%%%%%%%%%%%%%%%%%%%%%%%%%%%%%%
%%%%%%%%%%%%%%%%%%%%%%%%%%%%%%%%%%%%%%%%%%%%%%%%%%%%%%%%%%%%%%%%%%%%%%%%%%%%%%%%%
%%%%%%%%%%%%%%%%%%%%%%%%%%%%%%%%%%%%%%%%%%%%%%%%%%%%%%%%%%%%%%%%%%%%%%%%%%%%%%%%%

\subsection{Example 3: third order equation invariant under $\mathrm{SL}_x(2)$}

The differential equation: 
\begin{equation}\label{eq4}
\frac{1}{y'^2}\left(\frac{y'''}{y'}-\frac{3y''^2}{2y'^2}\right)=y
\end{equation}
is invariant under $\mathrm{SL}_x(2)$. Apparently, the general solution cannot be constructed in terms of elementary functions.

With the initial values 
\begin{equation}
y(0)=10,\quad y'(0)=-1,\quad y''(0)=-10
\end{equation}
the standard numerical methods stop at $x\approx 0.14$. A singularity (or a point with infinite derivative) is expected.

The invariant approach is obtained with the difference and $x$-lattice equations (we choose a uniform lattice, yielding $S=4$). The difference equation is a polynomial of third degree in $y_n$ and the search for real solutions becomes rather involved. However, it can be done and the results appear in the tables and graphics we present. The role of the other branches, which can contain complex values, is not well understood.

\begin{figure}[h]
\begin{center}
\includegraphics[scale=0.5]{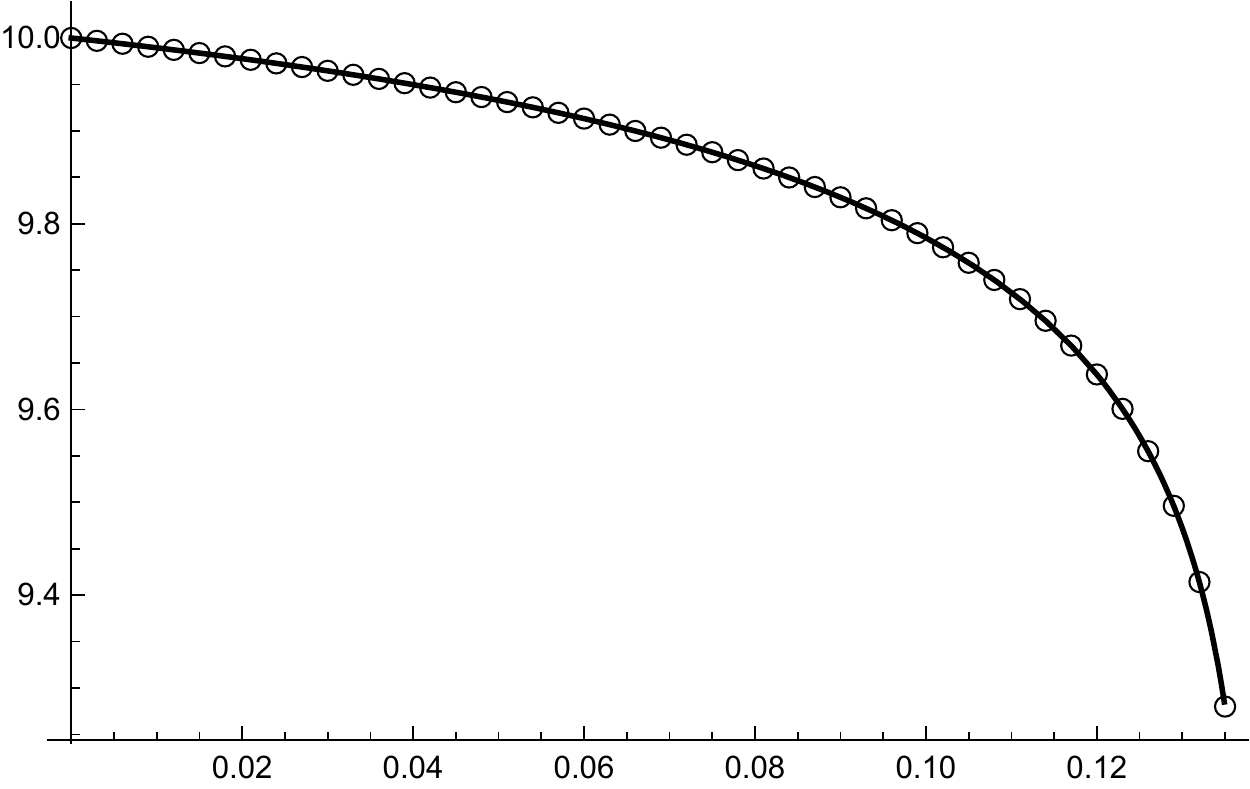}
\end{center}
\caption{Invariant approach (dots) versus standard approach (solid line) for equation (\ref{eq4}).\label{fig4}}
\end{figure}

If the initial conditions are taken from the approximate solution:
\begin{equation}
x_0=0.,\quad h=0.001, \quad y_0=10.,\quad y_1=9.9989, \quad y_2=9.9979
\end{equation}
the graphics of the invariant approach is given in Figure \ref{fig4}. Although the difference equation could be solved beyond the point $x=1.4$, $y_n$ becomes complex  (for the chosen branch). 

%%%%%%%%%%%%%%%%%%%%%%%%%%%%%%%%%%%%%%%%%%%%%%%%%%%%%%%%%%%%%%%%%%%%%%%%%%%%%%%
%%%%%%%%%%%%%%%%%%%%%%%%%%%%%%%%%%%%%%%%%%%%%%%%%%%%%%%%%%%%%%%%%%%%%%%%%%%%%%%
%%%%%%%%%%%%%%%%%%%%%%%%%%%%%%%%%%%%%%%%%%%%%%%%%%%%%%%%%%%%%%%%%%%%%%%%%%%%%%%
%%%%%%%%%%%%%%%%%%%%%%%%%%%%%%%%%%%%%%%%%%%%%%%%%%%%%%%%%%%%%%%%%%%%%%%%%%%%%%%

\subsection{Example 4: invariant equation under $\mathrm{SL}_x(2)\times\mathrm{SL}_y(2)$; a discrete exact solution}

We will consider differential equations invariant under the direct product group $\mathrm{SL}_x(2)\times \mathrm{SL}_y(2)$.

\begin{figure}[h]
\begin{center}
\includegraphics[scale=0.5]{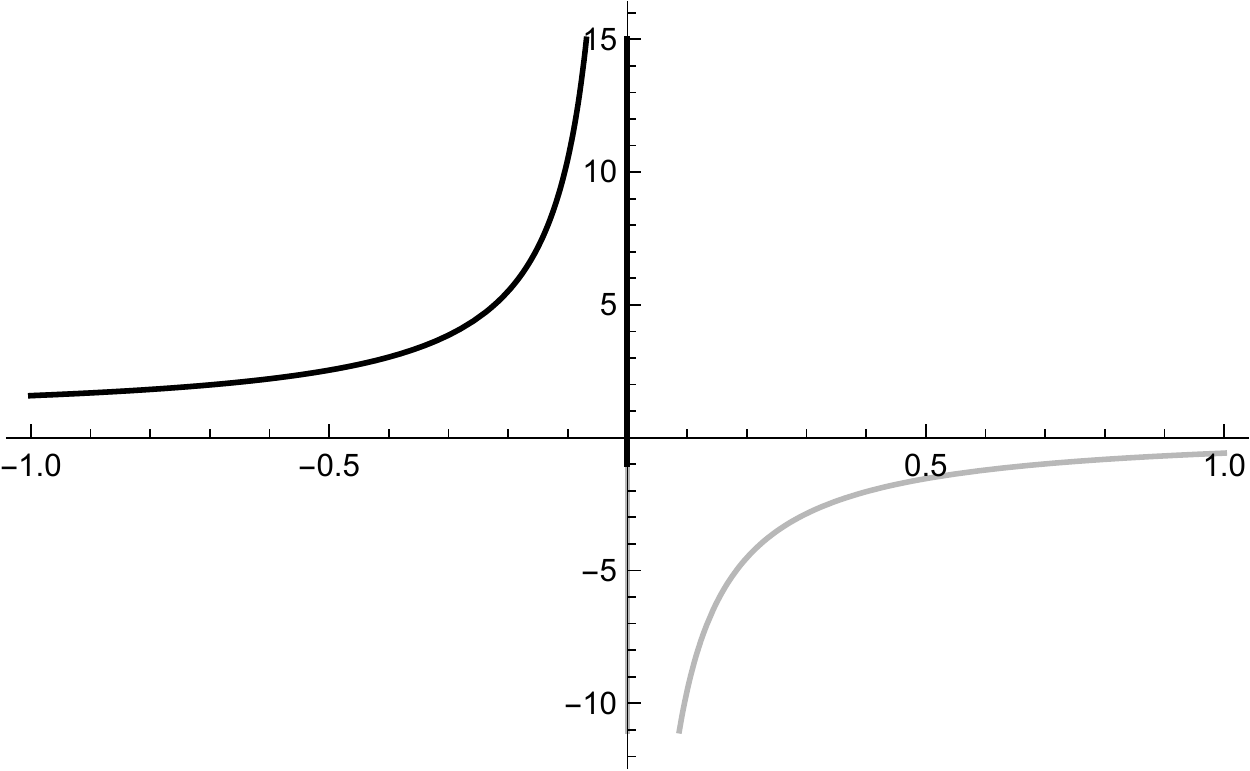}
\caption{The solution (\ref{sol5}) of equation (\ref{eq5}) obtained by a standard numerical method. Grey line, the exact solution.\label{fig5}}
\end{center}
\end{figure}

The equations are of the form
\begin{eqnarray}\label{eq5}
\fl H_5=c,\quad y^{(5)}= &\frac{1}{y'^3 \left(2 y'y''' -3 y''^2\right)} \left(\frac{5}{2} y'^4(y^{(4)})^2 -10   y'^3 y''y'''y^{(4)}+2 (c+4) y'^3y'''^3 \right.\nonumber \\ &\left.-\frac{9}{4} \left(c+\frac{2}{3}\right) \left(4  y'^2 y''^2y'''^2-6  y' y''^4y'''+3 y''^6\right)\right)
\end{eqnarray}
where $c$ is a constant.  The equation can be easily solved, although the general solution can adopt several equivalent forms. We will consider the particular solution:
\begin{equation}\label{sol5}
y(x)=\frac{1}{1-\re^x}, \quad c=0
\end{equation}
The values corresponding to this solution can be easily obtained by a standard numerical method with initial conditions:
\begin{equation}
y^{(k)}(x)=\left.\frac{\rd}{\rd x^k}\frac{1}{1-\re^x}\right|_{x=-1},\quad k=0,1,2,3,4
\end{equation}
until $x=0$ where the function has a singularity (see Figure \ref{fig5}, black curve).

\begin{figure}[h]
\begin{center}
\includegraphics[scale=0.5]{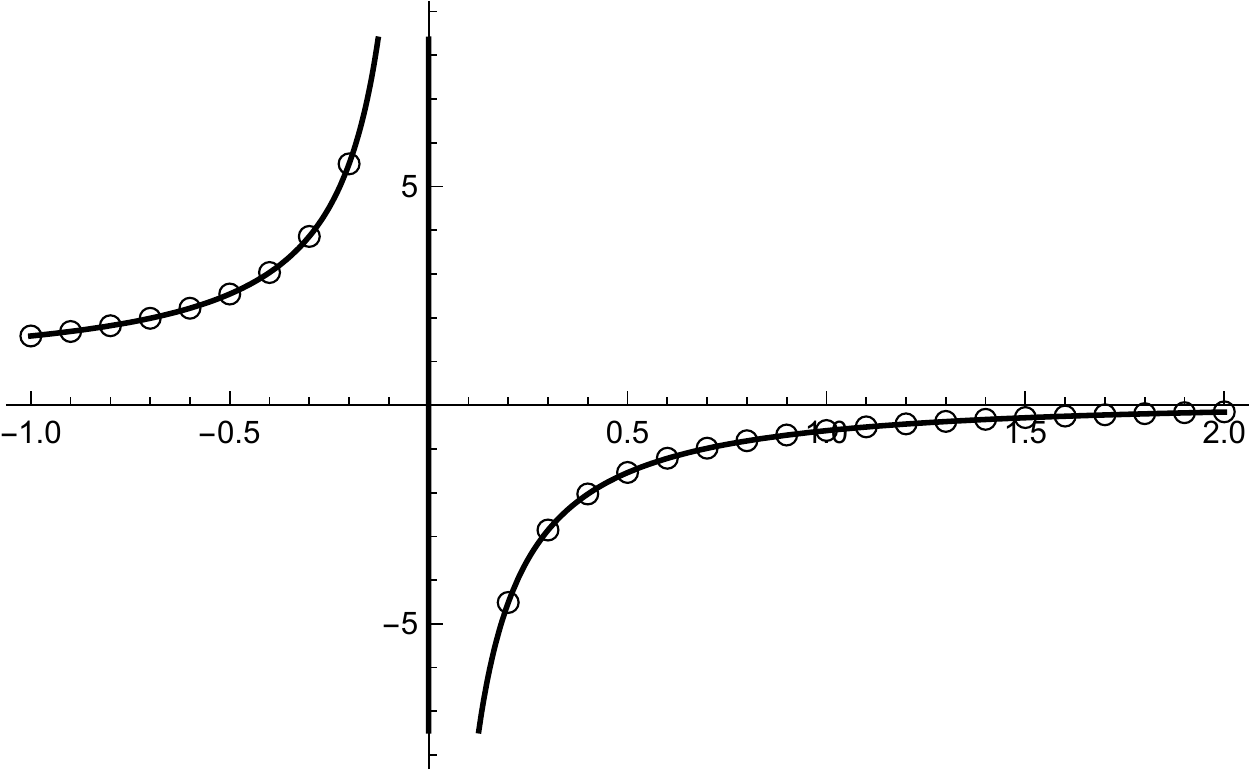}
\caption{The invariant approach (dots) versus the exact solution (\ref{sol5}) of equation (\ref{eq5})\label{fig6}}
\end{center}
\end{figure}

The invariant approach is obtained with the difference and $x$-lattice equations (we choose a uniform lattice, yielding $S=4$). 

\begin{equation}
\fl 3 R_4^2+(R_5-32) R_4+16 R_5+R_3 (R_4-5 R_5+16)=0,\quad 
x_n=x_0+ n h
\end{equation}
where
\begin{eqnarray}
\fl R_3=\frac{(y_{n-3}-y_{n-5}) (y_{n-2}-y_{n-4})}{(y_{n-4}-y_{n-5}) (y_{n-2}-y_{n-3})},\quad R_4= \frac{(y_{n-2}-y_{n-4}) (y_{n-1}-y_{n-3})}{(y_{n-3}-y_{n-4}) (y_{n-1}-y_{n-2})}\nonumber ,\\ 
\fl R_5= \frac{(y_{n-1}-y_{n-3}) (y_{n}-y_{n-2})}{(y_{n-2}-y_{n-3}) (y_{n}-y_{n-1})}
\end{eqnarray}
The difference equation is linear in $y_n$. 
The initial conditions are taken from the exact solution:
\begin{equation}
x_0=0,\quad h=0.1,\quad y_k=\frac{1}{1-\re^{x_k}},\quad k=0,1,2,3,4
\end{equation}
and the invariant solution is represented in Figure \ref{fig6}.

It turns out that the solution provided by the invariant approach is an exact solution:
\begin{equation}
y_n=\frac{1}{1-\re^{x_n}},\quad x_n=-1+nh
\end{equation}
This is easy to check. Compute $R_i$, for any four consecutive points, for instance $R_3$, when $y_n=\frac{1}{1-\re^{x_n}}$. We get (for a uniform lattice, $x_n=-1+hn$)
\begin{equation}
R_3=2+\re^{h}+\re^{-h}
\end{equation}
and the same expression for $R_4$ and $R_5$. Substituting in $H_5$ we get zero. This can be also observed from another point of view. The expression for $y_n$ is a solution of the equation:
\begin{equation}
R_k=\alpha
\end{equation}
and this provides a solution of the difference equation we are considering.

This observation allows to study the problem in the opposite direction. Since $R_k=\alpha$ is a solution of $H_5=0$, the equation $R_i=\alpha$ should provide solutions of the difference equation and approximation of solutions of the differential equation. These exact solutions have been computed in \cite{DK15} (for the case under study, see equation (5.24) of this reference).

%%%%%%%%%%%%%%%%%%%%%%%%%%%%%%%%%%%%%%%%%%%%%%%%%%%%%%%%%%%%%%%%%%%%%%%%%%%%%%%
%%%%%%%%%%%%%%%%%%%%%%%%%%%%%%%%%%%%%%%%%%%%%%%%%%%%%%%%%%%%%%%%%%%%%%%%%%%%%%%

\subsection{Example 5: invariant equation under $\mathrm{SL}_x(2)\times\mathrm{SL}_y(2)$}

Our final example is a discussion of the equation
\begin{equation}\label{eq6}
H_5=0
\end{equation}
and the particular solution:
\begin{equation}\label{sol6}
y(x)=\tan \frac{1}{x}
\end{equation}

\begin{figure}[h]
\begin{center}
\includegraphics[scale=0.5]{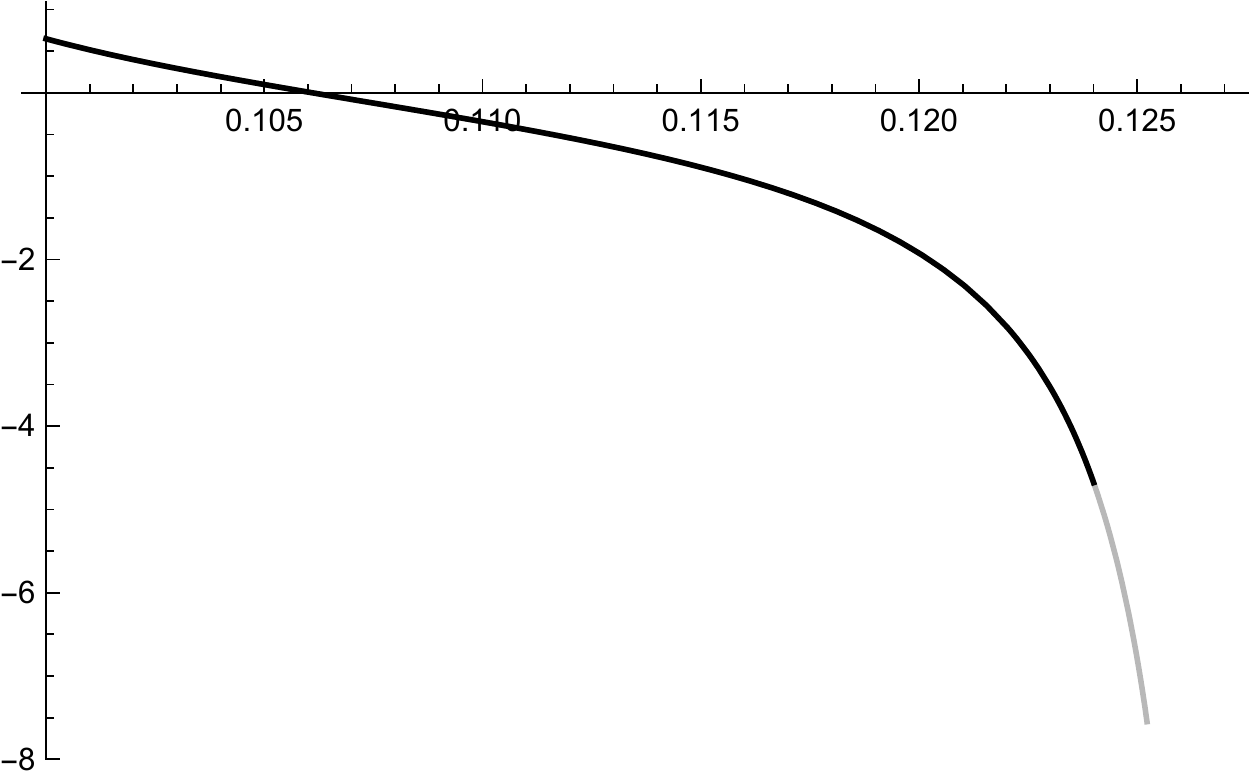}
\caption{Standard numerical approach versus the exact solution (\ref{sol6}) of equation (\ref{eq6}). The grey line corresponds to the exact solution.\label{fig7}}
\end{center}
\end{figure}

The numerical values of this solution can be easily obtained by a standard numerical method, although it cannot be prolonged beyond the singularities (the first one greater than  $x=0.1$ is located  at $x=2/(5\pi)$). We take an initial condition at $x_0=0.1$ with the values of the function and its derivatives computed using the exact solution (see Figure \ref{fig7}):
\begin{equation}
y(x_0)=\tan10,\quad y^{(k)}(x_0)=\left(\tan \frac{1}{x}\right)^{(k)}_{x=0.1},\quad k=1,\ldots,4
\end{equation}

The invariant approach is obtained with the difference and $x$-lattice equations (we choose a uniform lattice, yielding $S=4$):
\begin{equation}
\fl 3 R_4^2+(R_5-32) R_4+16 R_5+R_3 (R_4-5 R_5+16)=0,\quad
x_n=x_0+ n h
\end{equation}
and is linear in $y_n$. 
The initial conditions are also taken from the exact solution:
\begin{equation}
 x_0=0.1, \quad y_k=\tan\left(\frac{1}{0.1+kh}\right),\quad k=0,1,2,3,4
\end{equation}

The scheme is very sensitive to the step size and the fixed working precision. However, it is possible, using the invariant approach, to go beyond the singularities of the solution. We will just present a qualitative summary of results.

\begin{figure}[h]
\begin{center}
\includegraphics[scale=0.5]{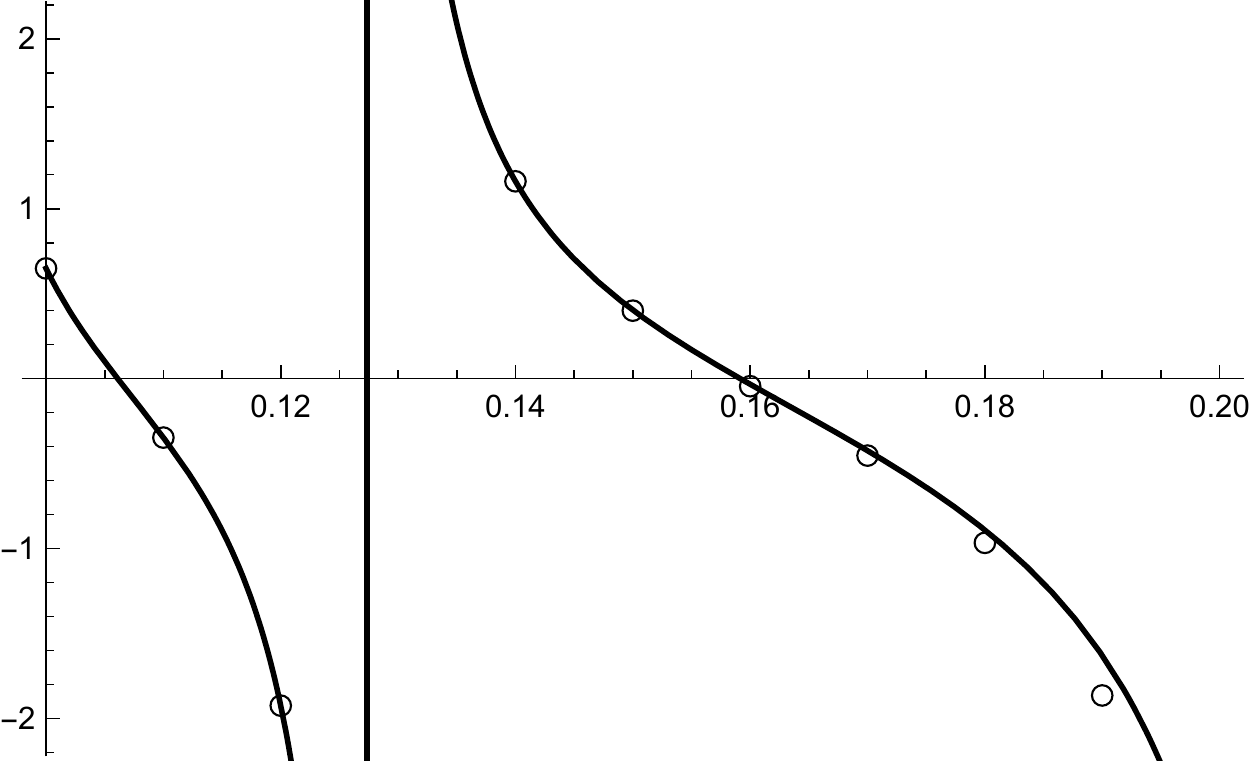}
\end{center}
\caption{Invariant approximation for solution (\ref{eq6}) of equation (\ref{sol6}). Step $h=0.01$\label{fig8}}
\end{figure}

\begin{figure}[h]
\begin{center}
\includegraphics[scale=0.5]{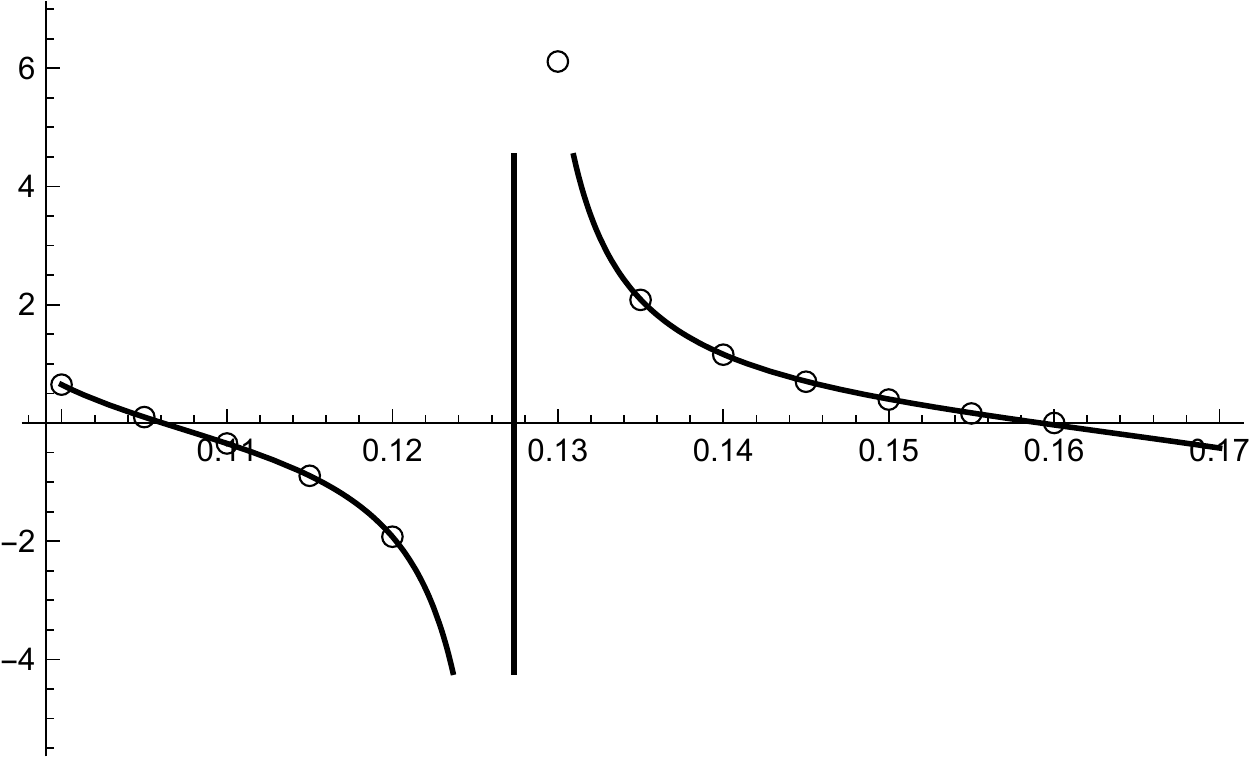}
\end{center}
\caption{Invariant approximation for solution (\ref{eq6}) of equation (\ref{sol6}). Step $h=0.005$\label{fig9}}
\end{figure}

\begin{figure}[h]
\begin{center}
\includegraphics[scale=0.5]{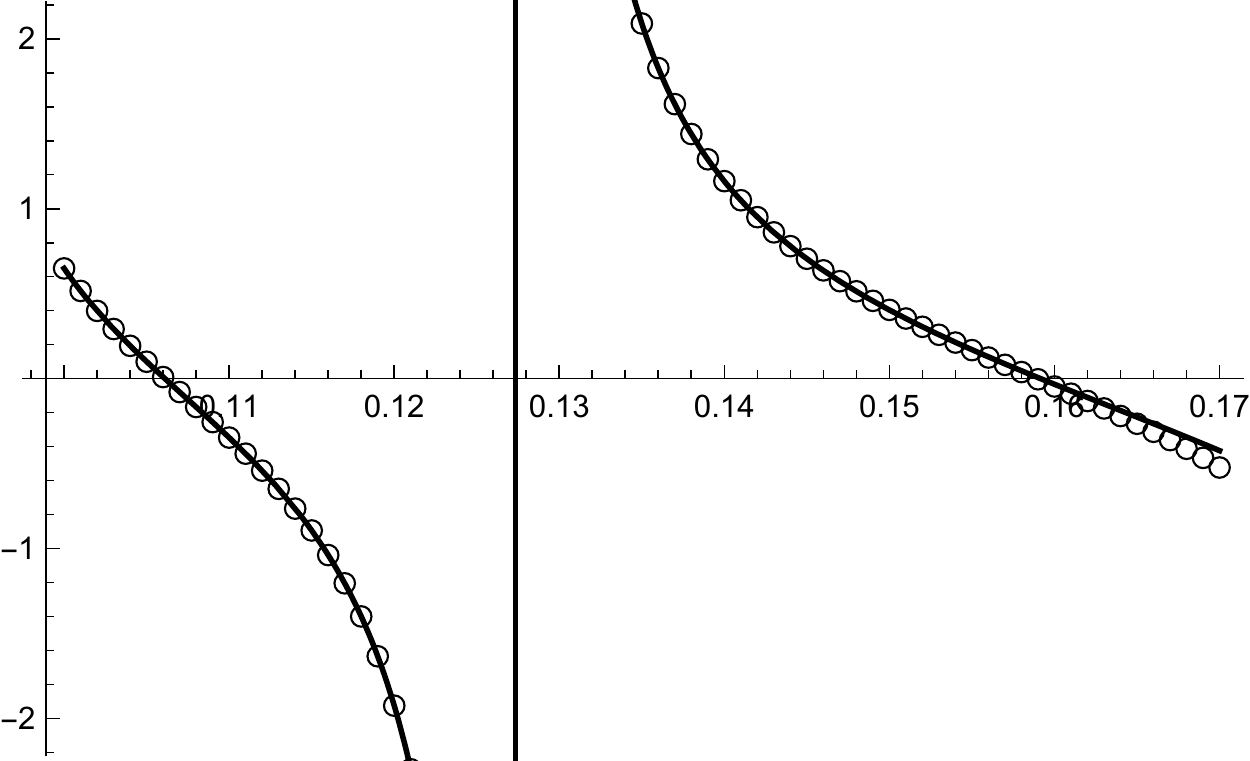}
\end{center}
\caption{Invariant approximation for solution (\ref{eq6}) of equation (\ref{sol6}). Step $h=0.001$\label{fig10}}
\end{figure}

The graphics in Figures \ref{fig8}, \ref{fig9} and \ref{fig10} represent the solution and the invariant discretization for $h=0.01$, $h=0.005$, and $h=0.001$, respectively.

Smaller steps provide a better approximation, although the roundoff errors prevent us (even at the cost of greater working precisions) to go beyond a certain point.

\section{Conclusions}
 The main theoretical results of this paper are contained in Section  4. We have shown that starting from the four-point difference invariants $R_i$ (\ref{invR}) and $S_i$ (\ref{invS}) of $\sly$ and $\slx$ we can construct  difference invariants of arbitrary order, for the 3 groups considered in this article. In the continuous limit they approach the corresponding differential invariants. Explicitly we go up to order $N=5$ and this provides invariant schemes for solving invariant ODEs of order up to five numerically. 

The numerical results are presented in Section 5. We consider several ODEs of order $N=3,4$ and $5$. The main features that emerge are the following:
\begin{enumerate}
\item Invariant numerical methods and standard methods provide very similar results for smooth solutions.
\item For solutions with singularities invariant methods provide significantly better results, specially close to singularities and beyond them.
 
\item In some cases the ``invariant numerical'' solutions are exact (see example 4 for a fifth order ODE. This is always true for first order ODEs \cite{RW04} and was already observed for some second order ones \cite{DK00a,DK00b}.
 
 \end{enumerate}
Since symmetries are an essential part of any physical problem  preserving them in a discretization is important in itself. This is true independently of whether invariant discretization improves numerical results.

An open question which merits further study is that of identifying equations and initial or boundary conditions for which invariant methods provide exact solutions. Work in this direction is currently in progress. Another line of research is related to the study of the several solutions arising from nonlinear discrete schemes (implicit schemes, where the highest point $y_n$ is not defined as a unique function of the previous points in the stencil).
% In the cases we have discussed in this work, different roots of higher degree polynomials provide solutions which do not necessarily  agree with the solution of the corresponding differential equation. Wether these functions are spurious, due to the discretization procedure, or have a meaning inside this theory, is a very interesting open question, presently under study.

\ack

MAR would like to thank Gabriel \'Alvarez for very fruitful discussions on the numerical computation in this research. Part of this work was completed during a series of visits to Centre de Recherches Math\'ematiques of the Universit\'e de Montr\'eal (Canada) and Universidad Complutense de Madrid (Spain). The authors wish to thank both Institutions for their support. RCS and MAR were supported by the Spanish Ministry of Science and Innovation under projects MTM2013-43820-P and  FIS2011-22566 respectively. The research of PW was partly supported by a research grant from NSERC of Canada and a Marie Sk{\l}odowska-Curie  fellowship from European Community. He thanks the Dipartimento di Matematica e Fisica di Roma Tre and specially Decio Levi for hospitality.

\end{document}